\documentstyle[floats,prb,aps,epsf,twocolumn]{revtex}
\begin{document}
\twocolumn[\hsize\textwidth\columnwidth\hsize\csname@twocolumnfalse\endcsname 
\draft
\title{Coherent effects in double-barrier Josephson junctions}
\author{A.\ Brinkman, A.\ A.\ Golubov}
\address{Department of Applied Physics, University of Twente,\\
P.O.Box 217, 7500 AE Enschede, The Netherlands}
\date{\today}
\maketitle

\begin{abstract}
The general solution for ballistic electronic transport through
double-barrier Josephson junctions is derived. We show the existence of a regime of
phase-coherent transport in which the supercurrent is proportional to the
single barrier transparency and the way in which this coherence is destroyed
for increasing interlayer thickness. The quasiparticle dc current at
arbitrary voltage is determined.
\end{abstract}


]

Phase-coherent electronic transport in mesoscopic structures between normal
(N) and superconducting (S) metals received considerable interest both in
experiments and in theory \cite{Been}. Particular interesting phenomena were
discovered in structures containing tunnel barriers (I). It is well known
that the subgap resistance of a ballistic SIN junction has a quadratic
dependence on the transparency of the interface\cite{BTK}, since Andreev
reflection is a two-particle process. Disorder in the normal region
enhances the Andreev current due to opening of some fraction of tunneling
channels, and the resistance has a linear dependence on the transparency.
This effect is known as reflectionless tunneling in SIN junctions (see \cite
{VZK,Nazar} and further references in \cite{Been}). Interestingly, the
opening of tunnel channels may be realized in a ballistic $NI_1NI_2S$
junction as well by placing a second tunnel barrier \cite{Mels}.

At the same time, a supercurrent $I_c$ in a tunnel $SIS$ contact depends
linearly on the barrier transparency since Cooper pairs tunnel coherently 
\cite{Jos,Likh}. Here we address the problem of universal features of
supercurrent flow in a double-barrier ballistic $SI_1S^{\prime }I_2S$
junction, where $S^{\prime }$ is a thin layer with critical temperature $%
T_{cs^{\prime }}<T_{cs}$. Coherent effects in such structures are also of
practical importance, since recent experiments demonstrated the possibility
of engineering Josephson junctions with desired properties using existing
multilayered techniques \cite{Ivan1,PTB}.

The supercurrent in a disordered double-barrier $SINIS$ junction was
calculated by Kupriyanov and Lukichev \cite{KL}, who considered the
interlayer in the dirty limit and $T_{cs^{\prime }}=0$. The coherent regime
was found in the limit of small interlayer thickness $d$, when supercurrent
is of the first order in the single barrier transmission coefficient $D$,
like in a $SIS$ junction. However in the limit of small $d$ the assumption
of the dirty limit is not justified since electronic scattering takes place
only at the interfaces. Theoretical work on ballistic $SINIS$ structures was
concentrated on studying resonant supercurrents in low-dimensional
structures \cite{GKL,Fur,Arne,Shum,GDK,Ivan2}.

In the present paper we study theoretically the universal features of charge
transport in a three-dimensional (3D) $SI_1S^{\prime }I_2S$ junction in the
clean limit. We demonstrate the existence of the coherent regime when the
supercurrent is averaged over the transmission resonances and is
proportional to $D$, whereas it becomes incoherent, of the order of $D^2$,
with increasing thickness, as expected for two uncorrelated sequential
tunneling processes. We study quantitatively the crossover between these two
regimes and the relation to the dirty limit results of Ref.\cite{KL}.
Further, we show that the coherent supercurrent can be exactly derived from
the distribution of transmission eigenvalues $\rho (D)\propto
D^{-3/2}(1-D)^{-1/2}$ known for a two-barrier $NI_1NI_2N$ contact\cite{Been}%
. Based on this distribution, we calculate the quasiparticle {\it dc}
current at arbitrary voltage, which shows signatures of multiple Andreev
reflections (MAR).

We consider a 3D ballistic $SI_1S^{\prime }I_2S$ contact, where S' is a thin
superconducting film with $T_{cs^{\prime }}<T_{cs}$ and the mean free path $%
l_{s^{\prime }}>>d$, $d$ is the interlayer thickness and $I_{1,2}$ are the
parallel atomically sharp interfaces with arbitrary transmission
coefficients. In the temperature Green's function method the supercurrent
density $J_s$ is expressed through the Fourier transform of the Green's
function $G\left( {\bf r},{\bf r}^{\prime }\right) $ over the transverse
coordinates 
\begin{eqnarray}
J_S(x) &=&\frac{i\hbar e}m\int \frac{d^2k_{\parallel }}{(2\pi )^2}%
T\sum_{\omega _n>0}  \nonumber \\
&&\lim_{x^{\prime }\rightarrow x}\left( \frac \partial {\partial x^{\prime }}%
-\frac \partial {\partial x}\right) G\left( x,x^{\prime };k_{\parallel
},\omega _n\right) ,  \label{Js}
\end{eqnarray}
where $x,x^{\prime }$ are the coordinates across the junction, $k_{\parallel
}$ is the wave-vector component in the junction plane, $\omega _n=(2n+1)\pi
T $ . The normal and the anomalous Green's functions $G(x,x^{\prime
}),F^{+}\left( x,x^{\prime }\right) $ obey the Gor'kov equations 
\begin{equation}
\left( 
\begin{array}{cc}
i\omega _n+H & \widetilde{\Delta }\left( x\right) \\ 
\widetilde{\Delta }^{*}\left( x\right) & i\omega _n-H
\end{array}
\right) \cdot \left( 
\begin{array}{c}
G \\ 
F^{+}
\end{array}
\right) =\left( 
\begin{array}{c}
\delta \left( x-x^{\prime }\right) \\ 
0
\end{array}
\right) ,  \label{Gor'kov}
\end{equation}
where $\widetilde{\Delta }=\Delta \exp (i\chi )$ is the spatially dependent
complex pair potential, $H=\frac{\hbar ^2}{2m}\frac{\partial ^2}{\partial x^2%
}+E_x-V\left( x\right) $, $E_x=$ $E_F-\hbar ^2k_{\parallel }^2/2m$ is the
electron kinetic energy across the junction, $E_F$ is the Fermi energy, $%
V(x)=W_1\delta \left( x\right) +W_2\delta \left( x-d\right) $ is the
interface potential, $W_{1,2}$ being the barrier strengths.

Let us choose the position $x^{\prime }$ within the interlayer, then the
solution of Eq. (\ref{Gor'kov}) in the superconducting electrodes ($\left|
x\right| >d/2)$ is given by a linear combination of plane waves $A(x^{\prime
})\exp (ik_xx)$. Substituting it into Eq. (\ref{Gor'kov}), we arrive the
dispersion relation for $k$ which yields four solutions: $k=\pm \sqrt{%
k_F^2-k_{\parallel }^2+2imE/\hbar ^2},k_{}^{*}=\pm \sqrt{k_F^2-k_{\parallel
}^2-2imE/\hbar ^2}$ with $E=\sqrt{\omega _n^2+\Delta ^2}$ and $%
k_F^2=2mE_F/\hbar ^2$. As a result the homogeneous solutions of Eq. (\ref
{Gor'kov}) at $x<-d/2$ has the form 
\begin{equation}
\left( 
\begin{array}{c}
G \\ 
F^{+}
\end{array}
\right) =a_{\pm }\left( 
\begin{array}{c}
1 \\ 
\beta
\end{array}
\right) e^{-ik(x\pm x^{\prime })}+b_{\pm }\left( 
\begin{array}{c}
-\beta ^{*} \\ 
1
\end{array}
\right) e^{ik^{*}(x\pm x^{\prime })},  \label{homog1}
\end{equation}
where $\beta =i\Delta ^{*}/(\omega _n+E)$. All four terms in Eq. (\ref
{homog1}) decay properly at $x=-\infty $. A similar solution holds for $%
x>-d/2$

\begin{equation}
\left( 
\begin{array}{c}
G \\ 
F^{+}
\end{array}
\right) =c_{\pm }\left( 
\begin{array}{c}
1 \\ 
\beta
\end{array}
\right) e^{ik(x\pm x^{\prime })}+d_{\pm }\left( 
\begin{array}{c}
-\beta ^{*} \\ 
1
\end{array}
\right) e^{-ik^{*}(x\pm x^{\prime })}.  \label{homog2}
\end{equation}

In the case of $T_{cs^{\prime }}=0$ the solution in the interlayer $\left|
x\right| <d/2$ consists of the homogeneous part, in which $\beta =0$ and all
terms of Eqs. (\ref{homog1}, \ref{homog2}) are present, and the particular
solution (source term): $G=$ $(m/i\hbar ^2k)e^{ik\left| x-x^{\prime }\right|
}$, $F=0$.

The solutions in all three regions are matched by the conditions of
continuity of the $G(x,x^{\prime })$ and $F(x,x^{\prime })$ at the
interfaces $x=\pm d/2$ and by the condition for the derivatives which
follows from the integration of Eq. (\ref{Gor'kov}) across the interface
barriers. This yields for $x=-d/2$ 
\begin{eqnarray}
&&0=\frac 1{2m_{S^{\prime }}}\frac{\partial G}{\partial x}\left( 
\begin{array}{c}
G \\ 
F^{+}
\end{array}
\right) _{x=d/2+0} \\
&&-\frac 1{2m_S}\frac \partial {\partial x}\left( 
\begin{array}{c}
G \\ 
F^{+}
\end{array}
\right) _{x=d/2-0}-W_1\left( 
\begin{array}{c}
G \\ 
F^{+}
\end{array}
\right) _{x=0},
\end{eqnarray}
and similar condition for $x=d/2.$ These boundary conditions provide the
required number of linear equations for the coefficients in the equations
for $G(x,x^{\prime })$ and $F(x,x^{\prime })$.

We can expand the wavevector in the interlayer as 
\begin{equation}
k_{s^{\prime }}=\sqrt{k_x^2+2im\sqrt{\omega _n^2+\left| \Delta _{s^{\prime
}}\right| ^2}/\hbar ^2}\simeq k_x+\frac i{2\xi _{s^{\prime }x}^{}},
\label{Eq.11}
\end{equation}
where $k_x=\sqrt{k_F^2-k_{\parallel }^2}=k_F\cos \theta $ is the transverse
component of the wave vector, $\xi _{s^{\prime }x}^{}=\xi _{s^{\prime }}^{}$ 
$/\cos \theta $, $\xi _{s^{\prime }}^{}=\hbar v_F/2\cos \theta \sqrt{\omega
_n^2+\Delta _{s^{\prime }}^2}$ is the coherence length and $\Delta
_{s^{\prime }}$ is the pair potential in the $S^{\prime }$ layer. In the $S$
electrodes we still keep $k_S=k_x$ with accuracy up to terms of the order of 
$\Delta /E_F.$ By matching the solutions (\ref{homog1}, \ref{homog2}) in all
three regions at the interfaces we derived an expression for the
supercurrent $J_s$, valid for arbitrary $d$ and $W_{1,2}$. Below we present
the solution for $d<\xi _{s^{\prime }}$and symmetric low-transparent
barriers $W_{1,2}=W,$ $(W/\hbar v_F)\gg 1$%
\begin{eqnarray}
J_s &=&\frac e\hbar \int \frac{d^2k_{\parallel }}{(2\pi )^2}T\sum_{\omega
\geq 0}  \nonumber \\
&&\frac{\Delta _s^2\sin \varphi +\Delta _s\Delta _{s^{\prime }}\sqrt{E_1/E_2}%
d/\xi _{s^{\prime }}^{}\widetilde{W}^2\sin \frac \varphi 2}{2\widetilde{W}%
^4E_1^2(\cosh d/\xi _{s^{\prime }x}^{}-\cos 2k_xd)+E_3^2}.  \label{eq:Js_gen}
\end{eqnarray}
Here $\varphi $ is the phase difference across the junction, $\Delta _s$ is
the pair potential in $S$, $E_1=\sqrt{\omega _n^2+\Delta _s^2}$, $E_2=\sqrt{%
\omega _n^2+\Delta _{s^{\prime }}^2}$, $E_3=\sqrt{\omega _n^2+\Delta
_s^2\cos ^2\varphi /2}$ and $\widetilde{W}=W/\hbar v_F$.

Eq. (\ref{eq:Js_gen}) is the main technical result of this paper and
describes the interplay between quasiparticle transmission (Breit-Wigner)
resonances and electron-hole (Andreev) resonances. Changing the phase space
in integration over $k_{\parallel }$ , one can apply Eq.(\ref{eq:Js_gen}) to
the problem of the supercurrent flow via transmission resonances in
low-dimensional contacts. In this case the results of \cite{Fur,Arne,Shum,GKL,GDK},
taken in relevant limits, are reproduced. Below we concentrate on the
interplay between coherent and incoherent regimes in the case of large $k_Fd$%
. It is instructive to discuss this crossover in terms of the width of
transmission resonances in a double-barrier junction, which in the
symmetrical case is given by $\Gamma =\hbar v_F\left\langle
xD(x)\right\rangle /2d=\hbar v_F/8d\widetilde{W}^2$, where $\left\langle
xD(x)\right\rangle $ is the angle-averaged transparency of a single barrier (%
$x=\cos \theta $).

{\it Coherent regime (broad resonances)}. As follows from Eq.(\ref
{eq:Js_gen}), the coherent regime takes place for a thin interlayer when the
transmission resonances are broad $\Gamma \gg \pi T_{cs}.$ The supercurrent
is then given by 
\begin{equation}
J_S=\frac e\hbar \int \frac{d^2k_{\parallel }}{(2\pi )^2}T\sum_\omega \frac{%
\Delta _s^2\sin \varphi }{E_1^2D_2^{-1}-\Delta _s^2\sin ^2\varphi /2},
\label{Js_coh}
\end{equation}
where $D_2$ is the transparency of a double barrier $NININ$ contact 
\begin{equation}
D_2^{-1}=1+\left( 2\widetilde{W}\cos k_F^{\prime }d+2\widetilde{W}^2\sin
k_F^{\prime }d\right) ^2,  \label{Trans}
\end{equation}
and has a resonant structure. Integration over the directions of $%
k_{\parallel }$ (over the resonances) yields the supercurrent

\begin{equation}
eJ_sR_N=2\pi T\sum_\omega \frac{\Delta _s^2\sin \varphi }{E_1E_3},
\label{KL}
\end{equation}
which does not depend on the properties of the interlayer and coincides with
the dirty-limit KL result \cite{KL}. Here $R_N^{-1}=e^2k_F^2\gamma /4\pi
^2\hbar $ is the normal state contact resistance per square, where $\gamma
=\left\langle xD(x)\right\rangle =1/4\widetilde{W}^2$. This expression can
be generalized to the asymmetric case: $eJ_sR_N=2\pi T\sum_\omega \left|
\Delta _S\right| ^2\sin \varphi /E_1E_3^{\prime }$ , with $E_3^{\prime }=%
\sqrt{\omega _n^2+\left| \Delta _s\right| ^2(\cos ^2\varphi /2+\gamma
_{-}^2\sin ^2\varphi /2)},$ $\gamma _{-}=(\gamma _1-\gamma _2)/(\gamma
_1+\gamma _2),$ $\gamma _{1,2}=\left\langle xD_{1,2}(x)\right\rangle $,
where $D_{1,2}$ are the individual barrier transparencies. In this case, $%
R_N^{-1}=e^2k_F^2\gamma _c/2\pi ^2\hbar $, where $\gamma _c=\gamma _1\gamma
_2/(\gamma _1+\gamma _2)$ which is the classical result known from \cite
{deJong}.

\begin{figure}
\par
\begin{center}
\mbox{\epsfxsize=0.9\hsize \epsffile{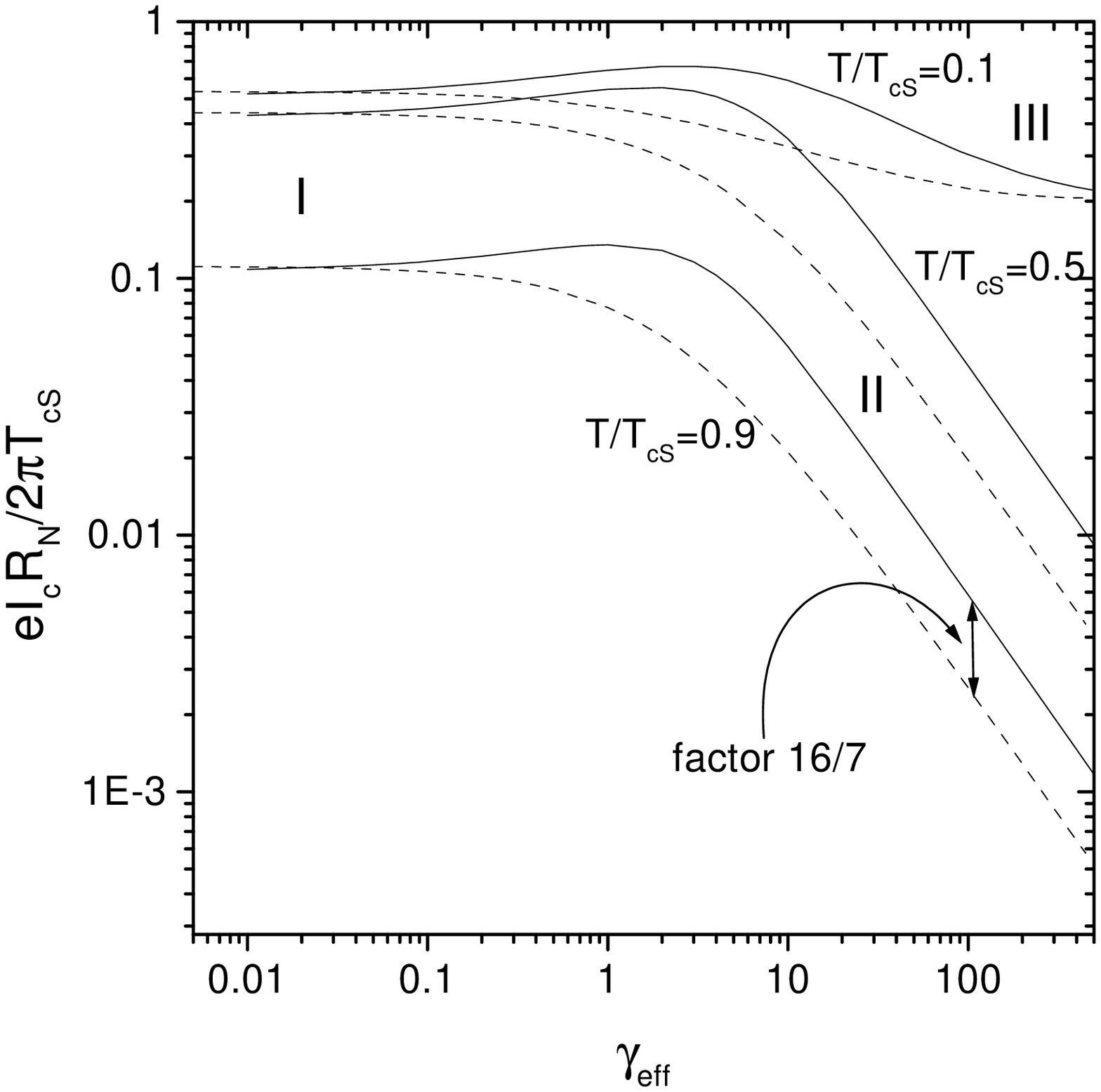}}
\end{center}
\caption
{$I_c R_N$ product in the model of clean interlayer (solid lines) 
and disordered interlayer (dashes lines).
The curves are plotted for the NB/Al case for $T_{cs}$ = 7.4 $T_{cs^{\prime}} $. }
\end{figure}

Expression (\ref{KL}) has been found by Kupriyanov and Lukichev \cite{KL} in
the case of a double-barrier junction with a dirty metal interlayer in the
limit of vanishingly small thickness. This fact shows that Eq. (\ref{KL}) is
a very general result.

For $T=0$ the maximum value of $eI_sR_N$ is achieved at $\varphi \approx
1.86 $ and exceeds the $eJ_cR_N$ value of $(\pi /2)\left| \Delta _s\right| $
for a tunnel $SIS$ contact. The reason is that in the coherent regime the
dominant contribution to $J_s$ comes from the transmission resonances, which
in the present case are broader than $\Delta .$ As a result the supercurrent
is of the first order in $\left\langle xD(x)\right\rangle $.

The supercurrent in the coherent regime has the spectral density 
\begin{equation}
%TCIMACRO{\func{Im} }
%BeginExpansion
\mathop{\rm Im}
%EndExpansion
J_s(\varepsilon )=\frac{\Delta _s^2\sin \varphi }{\sqrt{\Delta
_s^2-\varepsilon ^2}\sqrt{\varepsilon ^2-\Delta _s^2\cos ^2\varphi /2}},
\end{equation}
for $\Delta _s\cos \varphi /2<\varepsilon <\Delta _s$, while $%
%TCIMACRO{\func{Im}}
%BeginExpansion
\mathop{\rm Im}
%EndExpansion
J_S(\varepsilon )$ $=0$ for $\varepsilon <$ $\Delta _s\cos \varphi /2$ and $%
\varepsilon >\Delta _s$, i.e. the Andreev bound states in the energy range $%
\Delta _s\cos \varphi /2<\varepsilon <\Delta _s$ contribute to the
supercurrent.

{\it Incoherent regime (narrow resonances)} $\Gamma \ll \pi T_{cs}$.
With the increase of the interlayer thickness the coherent regime breaks
down due to the dephasing of the transmission resonances. After performing
the angle averaging in Eq.(\ref{eq:Js_gen}) the general expression for a
double-barrier junction becomes 
\begin{eqnarray}
eJ_sR_N &=&2\pi T\sum_{\omega _n}\frac{\Delta _s^2\sin \varphi }{E_1^2}%
\int_0^1\frac{4x^5dx}{\widetilde{W}^2\sqrt{a^2-1}}  \nonumber \\
&&+\frac{\Delta _s\Delta _{s^{\prime }}\sin \frac \varphi 2}{E_1E_2}\frac d{%
\xi _{s^{\prime }}^{}}\int_0^1\frac{8x^2dx}{\sqrt{a^2-1}},  \label{gen}
\end{eqnarray}
where $a=\cosh (d/\xi _{s^{\prime }}x)+\frac 1{2\widetilde{W}^4}(\Delta
_s^2\cos ^2\frac \varphi 2+\omega _n^2)/(\Delta _s^2+\omega _n^2)$. The pair
potential in S' is determined self-consistently with 
\begin{equation}
-\Delta _{s^{\prime }}\ln \frac T{T_{cs^{\prime }}}=2\pi T\sum_{\omega
_n}\left( \frac{\Delta _{s^{\prime }}}{\omega _n}-\left\langle F_{s^{\prime
}}\right\rangle \right) ,  \label{Self_cons}
\end{equation}
where $\left\langle F_{S^{\prime }}\right\rangle $ is the angle-averaged
anomalous Green's function, that is solved in the same way as the normal
Green's function G. 
\begin{equation}
\left\langle F_{s^{\prime }}\right\rangle =4\int_0^1(\frac{\Delta
_{s^{\prime }}}{E_2}\sinh (\frac d{\xi _{s^{\prime }}x})+\frac{2\Delta _s}{%
E_1}x^4\cos \frac \varphi 2)\frac{xdx}{\sqrt{a^2-1}}.  \label{Fs'}
\end{equation}

As seen from Eq.(\ref{gen}), the current-phase relation has two components, $%
\sin \varphi $ and $\sin \varphi /2$. The sign of the $\sin \varphi /2$
component is determined by the sign of the interlayer pair potential $\Delta
_{s^{\prime }}$, which is determined selfconsistently from Eq.(\ref
{Self_cons}) and depends on the sign of the electron-electron interaction in
an $S^{\prime }$ material. For an attractive interaction ($T_{cs^{\prime
}}>0 $) $\Delta _{s^{\prime }}$ does not vanish even at $T>T_{cs^{\prime }}$
and has a positive sign, while for a repulsive interaction $\Delta
_{s^{\prime }} $ becomes negative. Therefore the measurements of a
current-phase relation in a $SIS^{\prime }IS$ junction can be used for
measuring the sign of the electron-electron interaction in metallic films.

As is shown above, the critical current is controlled by a single
suppression parameter $\gamma _{eff}=\pi T_c/\Gamma =2\pi T_{cs}d/\hbar
v_F\left\langle xD(x)\right\rangle $. As follows from Eq.(\ref{gen}), for $%
T>T_{cs^{\prime }}$ and $\gamma _{eff}(\omega _n/\pi T_{cs})\gg 1$ the
supercurrent becomes of the order of $\left\langle xD(x)\right\rangle ^2$ as
expected for the incoherent tunneling in a double-barrier contact: 
\begin{equation}
eJ_sR_N=\frac{32\pi T}{7\gamma _{eff}}\sum_{\omega _n}\frac{\Delta _s^2\sin
\varphi }{E_1^2}.
\end{equation}
A numerical evaluation of Eq.(\ref{gen}) is shown in Fig. 1. The coherent
(I) and incoherent (II) regimes are indicated. Regime number III shows the
crossover to the series connection of two SIS' tunnel junctions.

{\it Transmission distribution, MAR.} The result for the coherent regime Eq.(%
\ref{KL}) can be derived from the transmission eigenvalue density $\rho
(D)=(G_N/\pi G_0)D^{-3/2}(1-D)^{-1/2}$ for two-barrier $NININ$ contacts \cite
{Been}, where $G_0=e^2/2\pi \hbar $. While deriving this distribution, an
assumption about a certain amount of impurity scattering in the interlayer
was made \cite{Been}; it can be shown that it also holds for the considered
case of a clean interlayer, provided that $k_Fd\gg 1$.

\begin{figure}
\par
\begin{center}
\mbox{\epsfxsize=0.9\hsize \epsffile{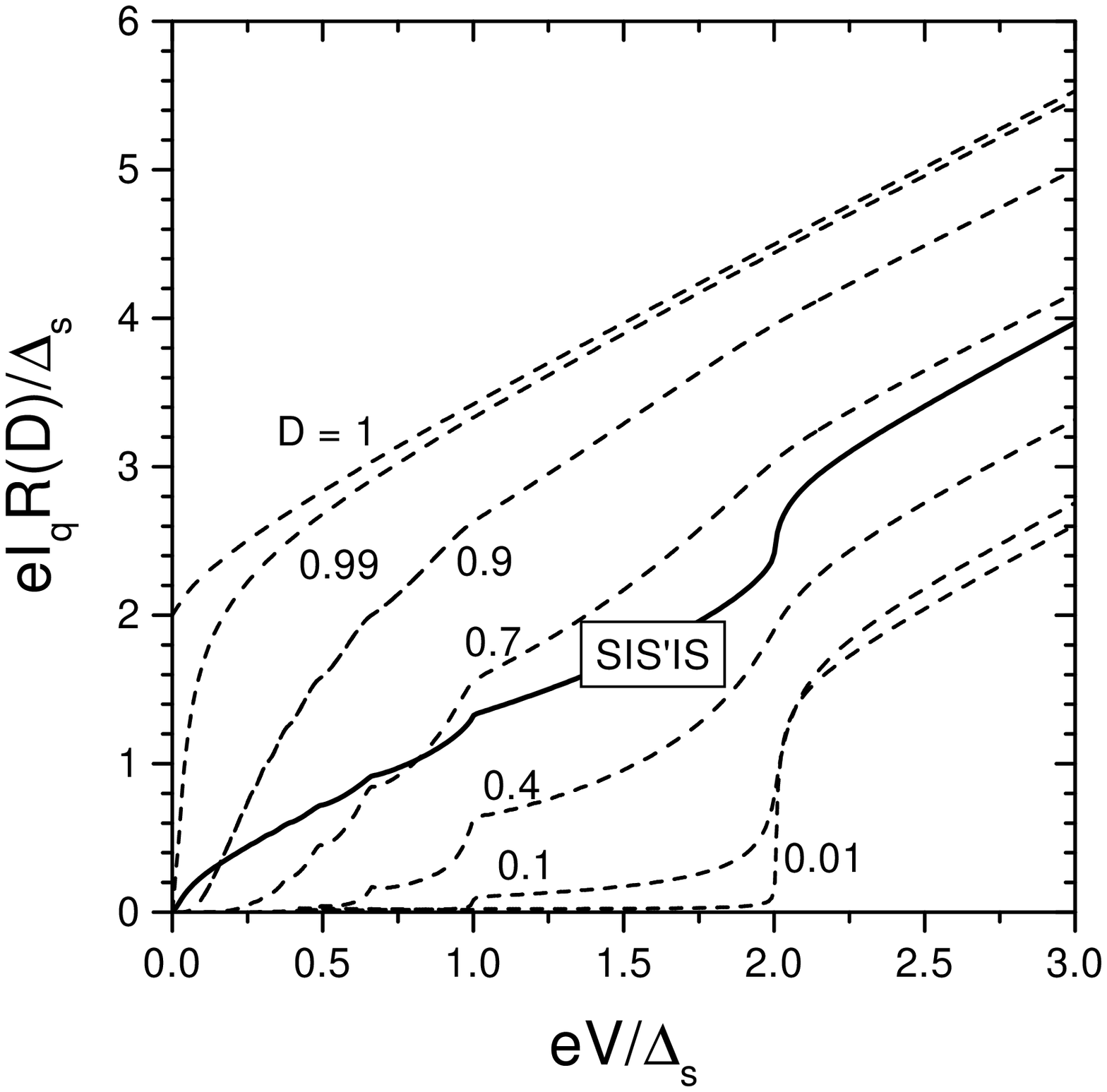}}
\end{center}
\caption{Dc current component in $SIS^{\prime }IS$ contact in the coherent 
regime, together with the single-channel curves for various transmissions D.}
\end{figure}

The derivation of Eq.(\ref{KL}) is then performed by the calculation of the
integral $\int_0^1$ $I_c(D)$ $\rho (D)dD$, where $I_c(D)$ is a supercurrent
per single ballistic channel, $I_c(D)=(e/2\hbar )\Delta _s^2D\sin \varphi
\tanh (E_B/2T)/E_B$ and $E_B=\Delta \sqrt{1-D\sin ^2\varphi /2}$, by the
residues of $\tanh (x)$ in a complex plane $z=(1/D-1)^{1/2}$. It yields
exactly the coherent supercurrent $eJ_sR_N=2\pi T\sum_{\omega \geq 0}\Delta
_s^2\sin \varphi /E_1E_3$. This proof also holds for the asymmetric case,
when the eigenvalue density is $\rho (D)=(G_N/\pi G_0)D^{-3/2}(D_{\max
}-D)^{-1/2}$ with $D_{\max }=4\gamma _1\gamma _2/(\gamma _1+\gamma _2)^2,$ $%
0<D<D_{\max .}$

The considerations above allow one to calculate the current under finite
voltage $V$ at the contact on the basis of the MAR (multiple Andreev
reflections) formalism \cite{Brat,Aver,Bard,Cue}. Below we discuss the dc
current component in a symmetric $SIS^{\prime }IS$ contact. The dc current
per single channel with transmisison D is given by \cite{Bard}

\begin{eqnarray}
\pi \hbar I_q(D,V)/e &=&eVD-\int d\varepsilon \tanh \frac \varepsilon {2T}%
(1-\left| a_0\right| ^2) \\
&&\times \left( 2%
%TCIMACRO{\func{Re} }
%BeginExpansion
\mathop{\rm Re}
%EndExpansion
(a_0A_0)+\sum_n(\left| A_n\right| ^2-\left| B_n\right| ^2)\right) , 
\nonumber
\end{eqnarray}
where $a_n=a(\varepsilon +neV)$ is the Andreev reflection amplitude: $%
a(\varepsilon )=(\varepsilon -sgn(\varepsilon )(\varepsilon ^2-\Delta
^2)^{1/2})$ for $\left| \varepsilon \right| >\Delta $ and $a(\varepsilon
)=(\varepsilon -i(\Delta ^2-\varepsilon ^2)^{1/2})$ for $\left| \varepsilon
\right| <\Delta $. The coefficients $A_n,B_n$ are given by the recurrency
relations which were derived in \cite{Bard} and can easily be solved
numerically.

The resulting total dc current in a symmetric $SIS^{\prime }IS$ contact is
given by the integration of the single-channel result $I_q(D,V)$ over the
eigenvalue density $I_q=\int_0^1I_q(D,V)\rho (D)dD$ where $\rho (D)=(G_N/\pi
G_0)D^{-3/2}(1-D)^{-1/2}$.

The results of numerical calculation at temperatures $T\ll T_{cs}$ are
presented in Fig.2. A few single-channel curves for various D are also shown
for comparison. The excess current $I_{ex}eR_N\simeq 1.05\Delta _s$ is
present at high bias $eV\gg \Delta _s$, while the subharmonic gap structure
at $eV=2\Delta _s/n$ due to MAR is present at lower voltages, despite
averaging over the channels. This dc current determines the amount of
dissipation in the junction. Note that the curve $SIS^{\prime }IS$ in Fig.2
is universal, i.e. independent of microscopic parameters such as electronic
mean free path or $T_c$ of the interlayer as long as $\gamma _{eff}<1$. This
universality breaks down with the increase of $\gamma _{eff}$, due to the
dephasing of the transmission resonances. The detailed theory applicable for
arbitrary $\gamma _{eff}$ is the subject of further study.

In conclusion, the general solution for the supercurrent in double-barrier
Josephson junctions is presented and the crossover from the coherent to the
incoherent regime for increasing interlayer thickness is studied in detail.
The parameter-free calculation of quasiparticle current is done in the
coherent regime. The results have implications for transport in Josephson
junctions and multilayers engineered with modern techniques.

The authors thank M.Yu. Kupriyanov, K.K. Likharev, Yu.V. Nazarov, I.P.
Nevirkovets and H. Rogalla for many useful and stimulating discussions.

\end{document}